\documentclass[twocolumn,superscriptaddress,pra]{revtex4}

\usepackage[dvips]{graphicx,color}
\usepackage{delarray}
\usepackage{ogonek}
\usepackage{amsmath}
\usepackage{bm}
\usepackage{mathrsfs}

\newcommand{\ket}[1]{| { #1} \rangle}
\newcommand{\bra}[1]{ \langle {#1} |}

\begin{document}
\title{Networking quantum networks with minimum cost aggregation}

\author{Koji Azuma}
\email{koji.azuma@ntt.com}
\affiliation{NTT Basic Research Laboratories, NTT Corporation, 3-1 Morinosato Wakamiya, Atsugi, Kanagawa 243-0198, Japan}
\affiliation{NTT Research Center for Theoretical Quantum Physics, NTT Corporation, 3-1 Morinosato-Wakamiya, Atsugi 243-0198, Japan}

\
\date{\today}

\begin{abstract}
A quantum internet holds promise for achieving distributed quantum sensing and large-scale quantum computer networks, as well as quantum communication among arbitrary clients all over the globe. 
The main building block is efficient distribution of entanglement---entangled bits (ebits)---between clients with fixed error, irrespective of their distance. In practice, this should be accomplished across multiple quantum networks, analogously to what the current Internet does in conventional communication.
However, existing quantum repeater schemes for efficient distribution of ebits over long distances work only on linear networks. Besides, there was no prescription on how to choose such linear networks from quantum networks with arbitrary topology. 
Here we present a practical recipe on how to aggregate quantum networks with arbitrary topology in order to give clients ebits with fixed error efficiently, regardless of their distance.
This recipe is based on combination of two new concepts of minimum cost aggregation and network concatenation.  
Our recipe forms the basis of designing a quantum internet protocol for networking self-organizing quantum networks to make a global-scale quantum internet.
\end{abstract}
\maketitle


\section{Introduction}

In a current internet, if a client, Alice, wants to communicate with another client, Bob, an internet protocol determines the path that the data follow to travel across multiple networks from Alice to Bob. In contrast, in the quantum version of an internet, called a quantum internet~\cite{K08,WEH18,ASEHJLT22}, there was no definitive recipe on how to transfer quantum data from Alice to Bob over multiple quantum networks. 
A reason for this is that in contrast to classical data, quantum data cannot be duplicated or amplified~\cite{WZ82,D82} and it is thus irreplaceable. Nevertheless, it is fragile in reality. Against this fragility, quantum error correction enables us to transmit quantum data even over noisy channels, like classical data in the current internet~\cite{KL96}. 
However, the allowed error thresholds to have the blessing of quantum error correcting codes are not so high. Hence, we cannot eliminate a risk of a catastrophic failure, e.g., because of an accidental error in the networks beyond the threshold, leading to dissipation of irreplaceable quantum data in the transmission.

On the other hand, if Alice and Bob share Bell pairs, called ebits,
quantum teleportation protocol enables Alice to send quantum data to Bob just by using local operations and classical communication (LOCC)~\cite{B93}. Since ebits are merely resource and replaceable and their existence is even testable by clients~\cite{BCPSW14}, there is no risk even if 
we set the goal of a quantum internet protocol to distribute ebits to clients. More than that, ebits are regarded as universal resource for clients in applications of a quantum internet~\cite{WEH18,ASEHJLT22}. For instance, clients can distil a secret key from ebits for information-theoretically secure communication~\cite{E91}. They can apply entangling operations between their local quantum devices~\cite{G99,CLP01,EJPP00} for distributed quantum computing, without meeting, just by consuming ebits. Moreover, a form of ebits can be used to extend the baseline between interferometric telescopes~\cite{GJC12}.
Besides, if ebits are shared among multiple clients, they can be transformed into arbitrary multipartite entangled states for multiparty protocols, such as Greenberger-Horne-Zeilinger states for distributed metrology~\cite{KKBJSYL14} and conference key agreement~\cite{AH09,CL07}. Therefore, the goal of a quantum internet protocol could be deemed as distribution of ebits to clients across given quantum networks, rather than direct transmission of quantum data via routing, 
implying that we cannot use protocols for a current internet as they are to control the quantum internet.

To distribute ebits efficiently between Alice and Bob over a linear network with repeater nodes between them, we can invoke the idea of quantum repeaters~\cite{B98}. However, this idea works only for linear networks and it does not tell us how we should choose such a linear network from a self-organizing large-scale quantum network with arbitrary topology. Recent proposals for aggregation of quantum repeaters have tried to answer this question~\cite{AK17, BAKE22,ABCE21}. However, they are unsatisfactory in practice, because they require point-to-point entanglement generation---the first step of the protocol---not only a) to suppress the error, depending on the whole size of the networks, but also b) to be run more than necessary. The former requirement a) implies that the error in the entanglement generation should be lowered to zero asymptotically, indicating that the proposals do not work in practical scenarios only with finite uses of the networks. The latter requirement b) indicates that they are costly to work, which gets worse as the sizes of the networks increase. Hence, there was no practical idea to control a quantum internet. 

In this paper, we present a practical recipe on how to aggregate quantum networks with arbitrary topology in order to give clients ebits with fixed error efficiently, regardless of their distance.
First, we introduce the concept of minimum cost aggregation which works over a quantum network. The cost here is a general notion like a price to pay for presenting ebits between two nodes in the network, which is minimized in the aggregation. This way, the aggregation eliminates the requirement b) above. 
Besides, the minimum cost aggregation is concatenated to enable arbitrary clients to have arbitrary long-distance communication with fixed error across multiple quantum networks, regardless of the overall size.
This ``network concatenation'' eliminates the requirement a) above. 
The idea of the concatenation is broader than conceptions of the nested entanglement purification protocol introduced by the original proposal for quantum repeaters~\cite{B98} and of concatenation of quantum error correcting codes~\cite{KL96}.
Our ideas form the basis to control a self-organizing large-scale quantum network in the future.

\section{Minimum cost aggregation}\label{sec:protocol}

We begin by introducing minimum cost aggregation in a quantum network (Fig.~1). The goal of this protocol is to present Bell pairs (or ebits) between two clients (or arbitrary two nodes) in a given quantum network, by utilizing the other nodes as repeater nodes. More precisely,
we associate a given quantum network with a graph $G=(V,E)$ with a set $V$ of vertices and a set $E$ of undirected edges, where vertices $x\in V$ correspond to quantum information processing nodes in the quantum network and each undirected edge $e=\{xy\}=\{yx\}\in E$ with $x,y\in V$ specifies quantum channels between different nodes $x$ and $y$ as a single quantum channel ${\cal N}_e$. We also associate two vertices $s(\in V)$ and $t(\in V)$ with the two clients and the other vertices  $r\in V\setminus \{s,t\}$ with repeater nodes. Besides, we assume that there is the maximum number $m^{\rm max}_e$ of uses of each channel ${\cal N}_e$, for instance, because of the availability of the channel for a given time interval.

An aggregation protocol starts by running a point-to-point entanglement generation scheme, perhaps equipped with quantum error correction or entanglement distillation, over quantum channel ${\cal N}_{e}$ for every $e\in E$, each of which provides a state $\hat{\rho}_{e}$ close to $f_{e} $ copies of a Bell pair $\ket{\Phi^+}_{e}:=(\ket{00}_e+\ket{11}_e)/\sqrt{2}$ for computation basis states $\{\ket{0},\ket{1}\}$, that is, $\| \hat{\rho}_{e} - \ket{\Phi^+}\bra{\Phi^+}_{e}^{\otimes {f_e}} \|_1 \le \delta_e$ with a given $\delta_e>0$ (in terms of the trace distance), by using the quantum channel ${\cal N}_e$ $ m_e(\le m_e^{\rm max})$ times and LOCC. 
As a result, we have a state $\hat{\rho}_{E}:=\bigotimes_{e\in E} \hat{\rho}_e$ close to Bell-pair network $\hat{\Phi}_E^f:=\bigotimes_{e\in E} \ket{\Phi^+}\bra{\Phi^+}^{\otimes  f_e}_e$.
Notice that $f_e$ is a monotonically non-decreasing function of $m_e$ for a fixed point-to-point entanglement generation scheme. Thus, if we write its maximum as $c_e:=f_e(m_e^{\rm max})$, we should have 
\begin{equation}
0\le f_e\le c_e\label{eq:cap}
\end{equation}
for any $m_e$ and $e \in E$.

In the existing protocol~\cite{AK17}, all the point-to-point entanglement generation schemes are assumed to use their own channel ${\cal N}_e$ the maximum number $m_e^{\rm max}$ of times, to establish a state close to a maximal Bell-pair network $\bigotimes_{e\in E} \ket{\Phi^+}\bra{\Phi^+}^{\otimes  c_e}_e$, rather than $\bigotimes_{e\in E} \ket{\Phi^+}\bra{\Phi^+}^{\otimes  f_e}_e$.
Then, it is transformed to Bell pairs between two clients $s$ and $t$ by performing entanglement swapping all over distinct $st$-paths composed of Bell pairs in the network $\bigotimes_{e\in E} \ket{\Phi^+}\bra{\Phi^+}^{\otimes  c_e}_e$.
According to Menger's theorem~\cite{M27,BM08}, there are 
\begin{equation}
C_{\{st\}} := \min_{V_{s;t}} \sum_{e\in \partial(V_{s;t})} c_e \label{eq:mincut}
\end{equation}
of such distinct paths, where $\partial(X)$ for a subset $X$ of $V$ is the set of undirected edges between nodes of $X$ and  nodes of $V\setminus X$ and the minimization is taken over all subsets $V_{s;t}$ of $V$ that include node $s$ (i.e., $s\in V_{s;t}$) but do not node $t$ (i.e., $t\notin V_{s;t}$ or $t \in V \setminus V_{s;t}$). 
As a result, the protocol provides the two clients $s$ and $t$ with a state close to $C_{\{st\}}$ copies of a Bell pair~\cite{AK17}.

\begin{figure}
    \includegraphics[width=70mm]{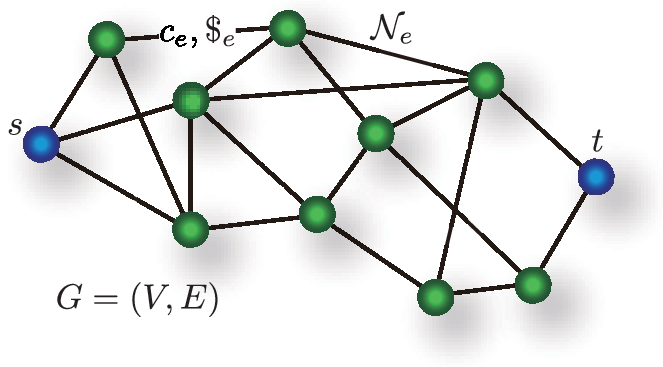}
    \caption{Quantum network associated with graph $G=(V,E)$. Vertices $x\in V$ correspond to quantum information processing nodes and each undirected edge $e=\{xy\}=\{ yx\} \in E$ specifies quantum channels between different nodes $x$ and $y$ as a single quantum channel ${\cal N}_e$. Two vertices $s(\in V)$ and $t(\in V)$ are held by two clients, and the other vertices $x\in V\setminus \{s,t\}$ are regarded as repeater nodes. If we run a point-to-point entanglement generation scheme by using quantum channel ${\cal N}_e$ the allowed maximum number $m_e^{\rm max}$ of times, it presents a state close to $c_e$ copies of a Bell pair $\ket{\Phi^+}_e$. $\$_e$ is a cost to produce an ebit on each edge $e$.}
    \label{fig:1}
\end{figure}

The goal of the current protocol includes providing the {\it same} number of Bell pairs to two clients $s$ and $t$ as the existing protocol~\cite{AK17}. However, in striking contrast, it is achieved with minimizing a total cost
\begin{equation}
\$^{f}:=\sum_{e\in E} \$_{e} f_e, \label{eq:cost1}
\end{equation}
where $\$_e(>0)$ is a cost to produce a unit of $f_e$, i.e., an ebit,  on each edge $e$. Notice that the cost $\$_e f_e$ for each edge $e \in E$ is proportional to $f_e$ and is thus monotonically non-decreasing for $m_e$, in general. As a result, our protocol does not require us to generate a state close to the maximal Bell-pair network $\bigotimes_{e\in E} \ket{\Phi^+}\bra{\Phi^+}^{\otimes  c_e}_e$ unless necessary, in contrast to the existing protocol~\cite{AK17}. 
More explicitly, for example, if we use entanglement generation schemes whose efficiencies $g_e:=  f_e/m_e$ can be deemed as constant and $\$_e=1/g_e$ is chosen as a cost, then $\$^{f}=\sum_{e\in E} f_e/g_e = \sum_{e\in E} m_e$, representing the total number of channel uses in the protocol. But, notice that this is not the only choice of the cost $\$_e$; for example, one could simply regard the cost $\$_e$ as a price of a Bell pair produced by an entanglement generation scheme over channel ${\cal N}_e$.

To establish the protocol, we associate the problem with the minimum cost flow, known in graph theory.
For that, we first notice an equivalence given by quantum teleportation~\cite{BAKE22}. The teleportation allows us to transmit $f_{\{xy\}}$ qubits in total by consuming $f_{\{xy\}}$ copies of a Bell pair, either from node $x$ to node $y$, i.e., to have a directed flow $f_{xy}$ with $f_{xy}=f_{\{xy\}}$, or from node $y$ to node $x$, i.e., to have a directed flow $f_{yx}$ with $f_{yx}=f_{\{xy\}}$. 
Conversely, if we can send $f_{xy}$ qubits from node $x$ to node $y$ faithfully, we can share $f_{\{xy\}}$ copies of a Bell pair between the nodes. 
Thanks to this equivalence, for the given graph $G(V,E)$, we can consider an induced digraph $D(V,A)$ which has the same set $V$ of vertices as the graph $G$ but has the set $A$ of directed edges $xy$ and $yx$ induced by all $\{xy\}\in E$.

In particular, Eq.~(\ref{eq:cap}) implies that the possible flows $f_{xy}(\ge 0)$ and $f_{yx}(\ge 0)$ of qubits, made by the quantum teleportation using $f_{\{xy\}}$ copies of a Bell pair, follow a capacity constraint,
\begin{equation}
0\le f_{xy} +f_{yx} \le c_{\{xy\}} \label{eq:cond1}
\end{equation}
for any $xy\in A$, $yx\in A$ and $\{xy\}\in E$. 
Besides, we require all the repeater nodes $r\in V \setminus \{s,t\}$ to make a flow $f$ of the quantum teleportation in a way satisfying a conservation condition, 
\begin{equation}
\sum_{a \in \partial^+(\{r\})} f_a =\sum_{a \in \partial^-(\{r\})} f_a, \label{eq:cond2}
\end{equation}
where $\partial^+(X)$ ($\partial^-(X)$)  for $X \subset V$ is the set of directed edges in $A$ whose tails belong to $X$ ($V\setminus X$) and whose heads belong to $V\setminus X$ ($X$). 
The net flow $F_{st}^f$ from node $s$ to node $t$ is described by
\begin{equation}
F_{st}^{f}:=\sum_{a \in \partial^+(\{s\})} f_a - \sum_{a \in \partial^-(\{s\})} f_a.
\end{equation}
If we maximize this net flow $F_{st}^f$ over flows $f$ on $A$ under the constraints (\ref{eq:cond1}) and (\ref{eq:cond2}), the maximum value is equal to the minimum cut $C_{\{st\}}$ of Eq.~(\ref{eq:mincut}), 
as stated in the max-flow min-cut theorem~\cite{F56,E56,BM08}.

On the other hand, since the cost for each edge $\{xy\}$ is regarded as
$\$_{\{xy\}} f_{\{xy\}} = \$_{\{xy \}} f_{xy}+\$_{\{xy \}} f_{yx}$,
the total cost of Eq.~(\ref{eq:cost1}) is rephrased as
\begin{equation}
\$^f=\sum_{a\in A} \$_{\{a\}} f_a. \label{eq:cost2}
\end{equation}
A solution of the linear program to minimize this total cost $\$^f$ over flows $f$ on $A$ under the net-flow constraint,
\begin{equation}
F_{st}^f=F^*_{\{st\}},
\end{equation}
for a given integer $F_{\{st\}}^*$ with
\begin{equation}
0 \le F_{\{st\}}^* \le C_{\{st\}}, \label{eq:F_{st}^*}
\end{equation}
as well as constraints (\ref{eq:cond1}) and (\ref{eq:cond2}), is called a minimum cost flow, denoted by $f^{*}$. The solution $f^*$ for the choice of $F_{\{st\}}^*=C_{\{st\}}$ is especially called a minimum-cost maximum flow. There are a variety of efficient algorithms to derive the minimum cost flows (see, e.g.,~\cite{AMO93}). Figure~2 presents an example of a solution of the minimum-cost maximum-flow problem. 

\begin{figure}
    \includegraphics[width=70mm]{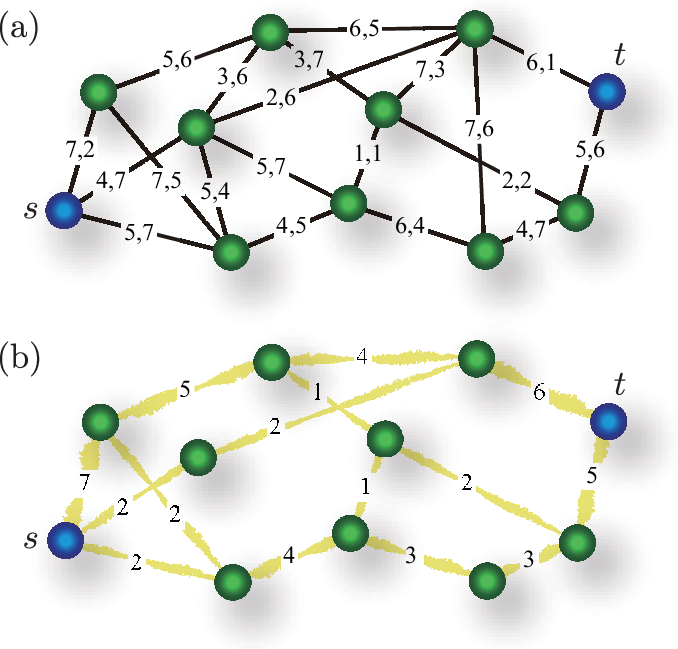}
    \caption{Minimum cost flow. As an example, suppose that two values on each edge $e$ in the graph (a) represent $c_e,\$_e$. Then, a solution $f^*$ of the minimum-cost maximum flow problem (that is, with assuming $F_{\{st\}}^*=C_{\{st\}}$)
    is described by values $f_e^*$ on edges $e$ in the graph (b). This solution shows $F_{\{st\}}^*=C_{\{st\}}=11$ and $\$^{f^*}=215$.}
    \label{fig:2}
\end{figure}

Now we can introduce our protocol to provide $F_{\{st\}}^* (\le C_{\{st\}})$ copies of a Bell pair $\ket{\Phi^+}_{\{st\}}$ to the clients $s$ and $t$ with minimum cost. 1)~The first step of this protocol is to derive a solution $f^{*}$ of the minimum-cost flow problem on $A$ for a given $F_{\{st\}}^*$. We define $f^*$ on $E$ as $f_{\{xy\}}^*:=f_{xy}^*+f_{yx}^*$ for any $xy\in A$ and $yx \in A$.
2)~Then, we run the point-to-point entanglement generation scheme over quantum channel ${\cal N}_{e}$ for every $e\in E$, to obtain a state $\hat{\rho}_e$ close to $f^*_{e}$ copies of a Bell pair $\ket{\Phi^+}_{e}$, that is, $\| \hat{\rho}_{e} - \ket{\Phi^+}\bra{\Phi^+}_{e}^{\otimes {f^*_e}} \|_1 \le \delta_e$ with a given $\delta_e>0$, by using the quantum channel ${\cal N}_e$ $ m_e^*(\le m_e^{\rm max})$ times and LOCC, where $m_e^*$ is defined as one satisfying\footnote{If there is no solution of $m_e$ satisfying $f^*_e=f_e(m_e)$, suppose that the smallest $m_e$ satisfying $f_e(m_e)> f_e^*$ is chosen as $m_e^*$.} $f_e^*=f_e(m_e^*)$.
3)~Finally, we perform an LOCC operation to make the quantum teleportation flow $f^*$ of qubits, so that it provides $F_{\{st\}}^*$ copies of a Bell pair $\ket{\Phi^+}_{\{st\}}$ between clients $s$ and $t$ if the initial state is the ideal Bell-pair network $\hat{\Phi}_{E^*}^{f^*}:=\bigotimes_{e \in E^*}\ket{\Phi^+}\bra{\Phi^+}^{\otimes {f^*_e}}_e$, where $E^*\subset E$ is the set of all undirected edges with $f_e^*>0$. 

If the LOCC operation in step 3) is a noiseless operation denoted by $\Lambda^{f^*}_{E^*}$ (satisfying $\Lambda_{E^*}^{f^*} ( \hat{\Phi}_{E^*}^{f^*}  )=\ket{\Phi^+}\bra{\Phi^+}^{\otimes {F_{\{st\}}^*}}_{\{st\}}$), the state $\hat{\omega}_{\{st\}}$ served to clients $s$ and $t$ after the application of the LOCC operation $\Lambda^{f^*}_{E^*}$ to the true initial state $\hat{\rho}_{E^*}:=\bigotimes_{e\in E^*}\hat{\rho}_e$ satisfies
\begin{align}
&\left\| \hat{\omega}_{\{st\}}- \ket{\Phi^+}\bra{\Phi^+}^{\otimes {F_{\{st\}}^*}}_{\{st\}} \right\|_1  =  \left\| \Lambda_{E^*}^{f^*} ( \hat{\rho}_{E^*} )-  \Lambda_{E^*}^{f^*} ( \hat{\Phi}_{E^*}^{f^*}  ) \right\|_1 \nonumber \\
&\le  \left\| \hat{\rho}_{E^*} -  \hat{\Phi}_{E^*}^{f^*} \right\|_1 \le  \sum_{e\in E^*} \delta_e=:\delta_{E^*}. \label{eq:error1}
\end{align}
Even if the LOCC operation in step 3) is merely a noisy operation $\Gamma^{f^*}_{E^*}$ rather than noiseless one $\Lambda^{f^*}_{E^*}$, the state $\hat{\omega}_{\{st\}}$ presented to clients $s$ and $t$ after the LOCC operation $\Gamma^{f^*}_{E^*}$ satisfies 
\begin{align}
&\left\| \hat{\omega}_{\{st\}}- \ket{\Phi^+}\bra{\Phi^+}^{\otimes {F_{\{st\}}^*}}_{\{st\}} \right\|_1 
= \left\| \Gamma_{E^*}^{f^*} (  \hat{\rho}_{E^*} )-  \Lambda_{E^*}^{f^*} ( \hat{\Phi}^{f^*}_{E^*} ) \right\|_1 \nonumber \\
& \le  \left\| \Gamma_{E^*}^{f^*} ( \hat{\rho}_{E^*} )-  \Gamma_{E^*}^{f^*} ( \hat{\Phi}_{E^*}^{f^*} ) \right\|_1 
 + \left\| \Gamma_{E^*}^{f^*} ( \hat{\Phi}_{E^*}^{f^*} ) -  \Lambda_{E^*}^{f^*} (  \hat{\Phi}_{E^*}^{f^*} ) \right\|_1 \nonumber \\
& \le  \delta_{E^*} + \epsilon_{E^*}, \label{eq:error2}
\end{align}
where $\epsilon_{E^*}:= \left\| \Gamma_{E^*}^{f^*} ( \hat{\Phi}_{E^*}^{f^*} ) -  \Lambda_{E^*}^{f^*} (  \hat{\Phi}_{E^*}^{f^*} ) \right\|_1$. 

On the other hand, since $f^*$ on $A$ is a minimum cost flow, this protocol merely needs the minimum cost $\$^{f^*} =\sum_{a\in A} \$_{\{a\}} f_a^*$. Therefore, with consuming this minimum cost $\$^{f^*}$, our protocol gives two clients $s$ and $t$ a state $\hat{\omega}_{\{st\}}$ $(\delta_{ E^*}+\epsilon_{E^*})$-close to $F_{\{st\}}^*$ copies of a Bell pair. Since $F^*_{\{st\}}$ can be chosen arbitrarily as long as it satisfies Eq.~(\ref{eq:F_{st}^*}), one might choose it so as to minimize a unit price of the ebits, $\$^{f^*}/F_{\{st\}}^*$ (or, so as to maximize the number of the ebits per cost, $F_{\{st\}}^*/\$^{f^*}$). 
Notice that the current protocol based on the choice of $F_{\{st\}}^*=C_{\{st\}}$ provides the same number of Bell pairs to the clients $s$ and $t$ as the existing protocol~\cite{AK17} (although our protocol works with minimum cost in contrast). As a result, for instance, similar to the existing protocol~\cite{AK17}, our protocol achieves the quantum and private capacities (per ``time'' $T$) of any quantum network composed of distillable channels~\cite{PLOB17} [whose relative entropies equal to the quantum capacities] 
 (see Appendix~\ref{sec:asympto} for the proof and Refs.~\cite{AK17,BAKE22,ABCE21,ASEHJLT22} for other implications).

\section{Network concatenation}

As shown in Sec.~\ref{sec:protocol}, 
the minimum cost aggregation in a given network $G=(V,E)$ gives clients $s(\in V)$ and $t(\in V)$ a state $\hat{\omega}_{\{st\}}$ $(\delta_{E^*}+\epsilon_{E^*})$-close to $F_{\{st\}}^*$ copies of a Bell pair $\ket{\Phi^+}_{\{st\}}$, by using the minimum cost $\$^{f^*}$. However, the errors $\delta_{E^*}$ and $\epsilon_{E^*}$ depend on the size of the network through the dependence on $|E^*|$, similar to the existing protocol~\cite{AK17} (although they are not larger than the errors in the existing protocol~\cite{AK17}). This implies that the point-to-point entanglement generation schemes, as well as LOCC operations for the quantum teleportation, should have asymptotically zero errors for arbitrary long-distance communication with fixed error. To overcome this point, we invoke the concept of concatenation~\cite{KL96} lying in the original proposal~\cite{B98} of quantum repeaters.
The main idea is to regard our aggregation to serve entanglement $\hat{\omega}_{\{st\}}$ to two clients $s$ and $t$ as an entanglement generation scheme on an undirected edge $\{st \} \in {\cal E}$ in a one level higher graph ${\cal G}=({\cal V},{\cal E})$, to which further minimum cost aggregation can be applied.

To clarify such a ``network concatenation'' (see Fig.~3),
suppose that each undirected edge $\varepsilon \in {\cal E}$ of a graph ${\cal G}=({\cal V},{\cal E})$ specifies a one level lower graph $G_\varepsilon=(V_\varepsilon,E_\varepsilon)$ corresponding to a quantum network in which our aggregation protocol can give the two ends of $\varepsilon$ a state $\hat{\omega}_\varepsilon$ $(\delta_{E_\varepsilon^*}+\epsilon_{E_\varepsilon^*})$-close to $F_{\varepsilon}^*$ copies of a Bell pair $\ket{\Phi^+}_{\varepsilon}$ by using the minimum cost $\$_\varepsilon^{f^*}$, where ${\cal V}$ is the set composed of all the two ends of edges $\varepsilon\in {\cal E}$. 
Here we assume that quantum networks $G_{\varepsilon}$ can work independently with each other and there is the maximum number $\mu_{\varepsilon}^{\rm max}$ of uses of each quantum network $G_\varepsilon$.
Besides, we assume that $\delta_{E_\varepsilon^*}+\epsilon_{E_\varepsilon^*}$ is small enough (at least, less than the threshold) to run a quantum error correcting code or an entanglement distillation protocol over every edge $\varepsilon\in {\cal E}$, which can convert $\mu_{\varepsilon}(\le \mu_{\varepsilon}^{\rm max})$ copies of the state $\hat{\omega}_{\varepsilon}$ to a state $\hat{\rho}_{\varepsilon}$ close to $\psi_{\varepsilon}$ copies of a Bell pair $\ket{\Phi^+}_{\varepsilon}$ with a given error $\delta_{\varepsilon}>0$, that is, $\| \hat{\rho}_{\varepsilon} - \ket{\Phi^+}\bra{\Phi^+}_{\varepsilon}^{\otimes {\psi_\varepsilon}} \|_1 \le \delta_\varepsilon$. 
These imply that we can obtain a state $\hat{\rho}_{{\cal E}}:=\bigotimes_{\varepsilon \in {\cal E}} \hat{\rho}_\varepsilon$ close to Bell-pair network $\hat{\Phi}_{\cal E}^{\psi}:=\bigotimes_{\varepsilon \in {\cal E}} \ket{\Phi^+}\bra{\Phi^+}^{\otimes  \psi_\varepsilon}_\varepsilon$ by using every quantum network $G_{\varepsilon}$ $ \mu_\varepsilon (\le \mu_\varepsilon^{\rm max})$ times (to have $\hat{\omega}_\epsilon^{\otimes \mu_\varepsilon}$) and by combining it with a quantum error correcting code or an entanglement distillation protocol. 
Similar to $f_e$ in Sec.~\ref{sec:protocol}, $\psi_\varepsilon$ is considered to be a monotonically non-decreasing function of $\mu_\varepsilon$ satisfying
\begin{equation}
0\le \psi_\varepsilon \le \theta_\varepsilon
\end{equation}
with $\theta_\varepsilon:=\psi_{\epsilon}(\mu_\varepsilon^{\rm max})$,
for any $\varepsilon \in {\cal E}$.

Besides, suppose that the cost to produce a unit of $\psi_{\varepsilon}$ over the graph $G_\varepsilon$ is written as $\pounds_\varepsilon$. As a result, the total cost of the aggregation over graph ${\cal G}$ is
\begin{equation}
\pounds^\psi:=\sum_{\varepsilon\in  {\cal E}} \pounds_\varepsilon \psi_\varepsilon.
\end{equation}
If the efficiency per cost  $\gamma_\varepsilon:=\psi_\varepsilon/(\mu_\varepsilon \$^{f^*}_{\varepsilon}) $ of the generation of the distilled entanglement $\hat{\rho}_\varepsilon$ over graph $G_\varepsilon$ can be assumed to be constant and $\pounds_{\varepsilon}=1/\gamma_\varepsilon$ is chosen as a cost, then $\pounds^\psi=\sum_{\varepsilon \in {\cal E}} \psi_\varepsilon/\gamma_{\epsilon}=\sum_{\varepsilon \in {\cal E}}  \mu_\varepsilon \$^{f^*}_{\varepsilon}$, representing the total cost in terms of the cost $\$^{f^*}_\varepsilon$ in the lower level of aggregation. Again, however, this is not the only choice of the cost $\pounds^\psi$.

\begin{figure}
    \includegraphics[width=70mm]{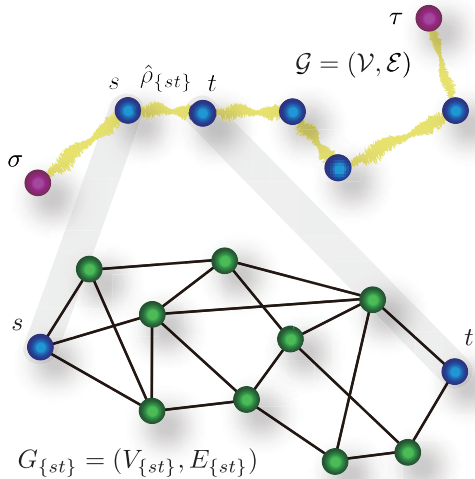}
    \caption{Network concatenation. For every edge $\varepsilon \in {\cal E}$, with the minimum cost aggregation, graph network $G_\varepsilon=(V_\varepsilon,E_\varepsilon) $ can be used to provide an entangled state $\hat{\rho}_{\varepsilon}$ close to Bell pairs for a one level higher graph network ${\cal G}=({\cal V},{\cal E})$. Based on these entanglement generations over edges $e\in {\cal E}$, we run the minimum cost aggregation over the graph ${\cal G}=({\cal V},{\cal E})$ to present ebits between nodes $\sigma \in {\cal V}$ and $\tau \in {\cal V}$. Further, these ebits themselves may be regarded as entanglement between nodes belonging to an even higher level graph network.}
    \label{fig:3}
\end{figure}

Then, we notice that the minimum cost aggregation over a graph $G=(V,E)$ in Sec.~\ref{sec:protocol} can work for two clients $\sigma\in {\cal V}$ and $\tau \in {\cal V}$ even over the one lever higher graph ${\cal G}=({\cal V},
{\cal E})$ with the following substitutions:
\begin{equation}
\begin{split}
G(V,E) \mapsto& {\cal G} ({\cal V},{\cal E}), \\
m_e (\le m_e^{\rm max} )\mapsto & \mu_{\varepsilon} (\le \mu_\varepsilon^{\rm max}),\\
f_e (\le c_e) \mapsto & \psi_{\varepsilon} (\le \theta_\varepsilon), \\
C_{\{st\}} \mapsto & \Theta_{\{\sigma \tau\}} , \\
\$_e \mapsto & \pounds_{\varepsilon} ,
\end{split}
\end{equation}
where 
\begin{equation}
\Theta_{\{\sigma \tau\}} := \min_{{\cal V}_{\sigma;\tau}} \sum_{\varepsilon\in \partial ({\cal V}_{\sigma;\tau})} \theta_\varepsilon
\end{equation}
minimized over subsets ${\cal V}_{\sigma;\tau}$ of ${\cal V}$ that include node $\sigma$ (i.e., $\sigma \in {\cal V}_{\sigma;\tau}$) but do not node $\tau$ (i.e., $\tau \notin {\cal V}_{\sigma;\tau}$ or $\tau \in {\cal V} \setminus {\cal V}_{\sigma;\tau}$).
In particular, for any two clients $\sigma\in {\cal V}$ and $\tau\in {\cal V}$ and for a given integer ${\Psi}_{\{\sigma \tau\}}^*$ with
\begin{equation}
0\le \Psi_{\{\sigma \tau\}}^* \le \Theta_{\{\sigma \tau\}},
\end{equation}
based on a derived minimum cost flow $\psi^*$, the aggregation over the one level higher network ${\cal G}$ gives two clients $\sigma$ and $\tau$ state $\hat{\omega}_{\{\sigma \tau\}}$ $(\delta_{{\cal E}^*} + \epsilon_{{\cal E}^*})$-close to $\Psi_{\{\sigma \tau\}}^*$ copies of a Bell pair $\ket{\Phi^+}_{\{\sigma \tau\}}$ by using the minimum cost $\pounds^{\psi^*}$, where ${\cal E}^*$ represents edges $\varepsilon \in {\cal E}$ with $\psi_\varepsilon^*>0$ and
\begin{align}
\delta_{{\cal E}^*}:= & \sum_{\varepsilon \in {\cal E}^*} \delta_\varepsilon, \\
\epsilon_{{\cal E}^*}:= & \left\| \Gamma_{{\cal E}^*}^{\psi^*} ( \hat{\Phi}_{{\cal E}^*}^{\psi^*} ) -  \Lambda_{{\cal E}^*}^{\psi^*} (  \hat{\Phi}_{{\cal E}^*}^{\psi^*} ) \right\|_1
\end{align}
are defined analogously to Eqs.~(\ref{eq:error1}) and (\ref{eq:error2}). Since $\delta_\varepsilon$ could be chosen arbitrarily (as long as there is a quantum error correcting code or an entanglement distillation protocol as required), the error $\delta_{{\cal E}^*} + \epsilon_{{\cal E}^*}$ depends only on the size of the current graph ${\cal G}$, independent of those of one level lower graphs $G_\varepsilon$. This is an advantage of concatenating aggregation.

Repeatedly, if the error $\delta_{{\cal E}^*} + \epsilon_{{\cal E}^*}$ of the current aggregation is small enough (at least, less than a threshold) to perform a quantum error correcting code or an entanglement distillation protocol (which can transform copies of the state $\hat{\omega}_{\{\sigma \tau\}}$ to a state close to fewer copies of a Bell pair $\ket{\Phi^+}_{\{\sigma \tau\}}$), we can consider further aggregation over an even higher level of a graph network by regarding the current aggregation as an entanglement generation scheme on an undirected edge $\{\sigma \tau\}$ in the higher-level graph. Therefore, by properly concatenating the aggregation, we can provide copies of a Bell pair with a fixed error to any two clients across quantum networks with arbitrary topology, independently of the whole size. 

Notice that the idea of the network concatenation includes conceptions not only of the nested entanglement purification protocol introduced by the original proposal for quantum repeaters~\cite{B98}, but also of the concatenation of quantum error correcting codes~\cite{KL96} (thanks to an equivalence between quantum error correction and entanglement distillation~\cite{BDSW96}). Indeed, this reduction is possible just by applying our network concatenation idea to linear networks, sequentially.

\section{Discussion}

In the future, it would be natural that the minimum cost aggregation is applied to quantum subnetworks, independently, as subnetworks would be governed by different network holders, such as individuals, public, institutes, universities, companies, and governments, like the current internet. 
Then, the idea of the network concatenation would play the role analogous to the current Internet, that is, to connect subnetworks to have ``a network of networks,'' although they differ in what they serve: the former distributes resource in the form of correlation, while the latter transmits data itself. As a result, a network design for the quantum internet could be substantially different from classical one.

We have only just begun to grasp a possible form of the future quantum internet: for example, our recipe would be specified more by considering available quantum error correcting codes, entanglement distillation protocols and effects of time lags of the use of classical communication between distant sites (as explored in the context of conventional quantum repeater protocols~\cite{RPL08,J09,M10,M12,AML16,FWHLMH10,ATL15}), as well as available physical devices. 
Following these, efficient algorithms to control a real quantum internet would also appear. 
Since the cost introduced in this paper can be associated with the price of a Bell pair, our theory would also be the basis of pricing quantum communication services served by a quantum internet, that is, the economics of a quantum internet, which is related to one of most important open questions raised by Lo in Ref.~\cite{ASEHJLT22}.

\section*{ACKNOWLEDGMENTS}
We thank Stefan B\"auml, David Elkouss, Toshimori Honjo, Hoi-Kwong Lo, William J. Munro, and Kiyoshi Tamaki for helpful discussion. 
We acknowledge the support, in part, from Moonshot R\&D, JST JPMJMS2061, from CREST, JST JP-MJCR1671, and from JSPS KAKENHI 21H05183 JP.

\appendix

\section{ASYMPTOTIC LIMIT}\label{sec:asympto}

Let us derive the ultimate performance of our protocol by taking an asymptotic limit. For that, we assume that all the local operations are
noiseless and introduce a parameter $T$ like ``time,'' which is arbitrary as long as rates $r_e:=m_e^{\rm max}/T$ are constant. Then, 
for the fixed rate $r_e>0$, $m_e^{\rm max}\to \infty$ in the limit of $T\to \infty$, 
for which, by optimizing the point-to-point entanglement generation schemes used in our protocol, $f_e(m_e^{\rm max})/m_e^{\rm max}=c_e/m_e^{\rm max}$ could reach the quantum capacity $Q^\leftrightarrow({\cal N}_e)$ of channel ${\cal N}_e$ assisted by unlimited forward and backward classical communication and the error $\delta_e$ follows $\delta_e \to 0$. Then, the number $F^*_{\{st\}}$ of ebits served to clients $s$ and $t$ per ``time'' $T$ in the asymptotic limit of $T\to \infty$, i.e., the asymptotic rate $R:=\lim_{T\to \infty} F^*_{\{st\}}/T$ is upper bounded as
\begin{align}
R\le& \lim_{T\to \infty} \frac{C_{\{st\}}}{T} 
= \lim_{T\to \infty} \min_{V_{s;t}} \sum_{e\in \partial(V_{s;t})} \frac{m_e^{\rm max}}{T} \frac{c_e}{m_e^{\rm max}}  \nonumber \\
\le &  \min_{V_{s;t}} \sum_{e\in \partial(V_{s;t})} r_e Q^\leftrightarrow({\cal N}_e).
\end{align}
Clearly, this upper bound is achievable by our protocol with $F^*_{\{st\}}=C_{\{st\}}$ using the optimal point-to-point entanglement generation schemes, in which the served ebits are asymptotically noiseless according to Eq.~(\ref{eq:error1}). 
Therefore, our protocol provides $\min_{V_{s;t}} \sum_{e\in \partial(V_{s;t})} r_e Q^\leftrightarrow({\cal N}_e)$ ebits to the clients $s$ and $t$ per ``time'' $T$ asymptotically, which is the same performance as the existing protocol~\cite{AK17} although the consumed cost in our protocol is minimum in striking contrast. 
As a result, for instance, similar to the existing protocol~\cite{AK17}, our current protocol achieves the quantum and private capacities (per ``time'' $T$) of any quantum network composed of distillable channels~\cite{PLOB17} [whose relative entropies equal to the quantum capacities] (see Refs.~\cite{AK17,BAKE22,ABCE21,ASEHJLT22} for other implications).

\end{document}